\documentclass{aa}  
\pdfoutput=1

\usepackage{graphicx}
\usepackage{color}

\usepackage{subcaption}
\usepackage{longtable}

\usepackage{txfonts}
\usepackage{color}

\begin{document}

\title{Bar quenching: Evidence from star-formation-rate indicators}

\author{K. George\inst{1}\fnmsep\thanks{koshyastro@gmail.com}, S. Subramanian\inst{2}}

\institute{Faculty of Physics, Ludwig-Maximilians-Universit{\"a}t, Scheinerstr. 1, Munich, 81679, Germany  \and Indian Institute of Astrophysics, Koramangala II Block, Bangalore, India}

  \abstract{The central regions of star-forming barred spiral galaxies can be devoid of star formation because of the redistribution of gas along the length of the bar. However, there can be gas outside the length of the bar that can  host star formation. We study a sample of barred disc galaxies in the local Universe with an aim to discriminate between centrally quenched and globally quenched galaxies based on their positions on star-formation-rate(SFR)--stellar mass plots and to find a connection between the SFR of quenched galaxies and the length of their bar. We classified barred galaxies as centrally quenched and globally quenched based on their position on SFR--stellar mass plots, with SFRs derived from H$\alpha$ flux and spectral energy distribution fits on combined ultraviolet and optical flux. We selected galaxies as passive based on the distance from the main sequence relation. From a total 2514 barred galaxies studied here, we present 651 with suppressed star formation in their central region but hosting star formation outside. We also find a possible correlation between bar length and SFR for the galaxies that are fully quenched because of the stellar bar.}


\maketitle
%

\section{Introduction}

Observations of barred disc galaxies in the local Universe reveal that the bar region, the annular region between the central sub-kiloparsec(sub-kpc)-scale nuclear star-forming region and the ends of the stellar bar, is devoid of recent star formation \citep{James_2015,James_2016,James_2018,George_2019a,George_2020}. This could be due to redistribution of cold gas along the length of the stellar bar. The bar-induced torque drives gas inflows to the center of galaxy. This enhances the star formation in the central sub-kpc-scale nuclear region and depletes the bar region of the fuel required for further star formation (\citealt{Combes_1985,Spinoso_2017,George_2019a,Newnham_2020,George_2020}). Another possible mechanism is related to the bar-induced shocks and shear, which increase the turbulence of the gas in the bar region and in turn stabilise the gas against collapse, leading to inhibition of star formation (\citealt{Tubbs_1982}; \citealt{Reynaud_1998}; \citealt{Verley_2007}; \citealt{Haywood_2016}; \citealt{Khoperskov_2018}). Thus, the action of the stellar bar in barred disc galaxies can suppress star formation in the bar region, a process known as bar quenching \citep{Tubbs_1982,Reynaud_1998,Masters_2010,Masters_2012,Cheung_2013,Renaud_2013,Gavazzi_2015,Hakobyan_2016,James_2016,Cervantes_2017,Spinoso_2017,Khoperskov_2018,James_2018,Kruk_2018,George_2019a,Donohoe-Keyes_2019, Newnham_2020, Rosas-Guevara_2020,Krishnarao_2020, Diaz_2020, Fraser-McKelvie_2020, George_2020, Zhang_2021}. Bar quenching could be one of the primary mechanisms that facilitate the global quenching of star formation in massive barred galaxies, transforming them into passive galaxies \citep{Man_2018}.\\

The effect of bar quenching is the formation of a cavity region along the length of bar with no star formation. This bar region is also referred to as the star formation desert \citep{James_2009}. As the suppression of star formation happens in the region covered by the length of the bar, it is natural to expect the impact of bar quenching to be more significant in galaxies with a longer bar. Though other quenching mechanisms, such as cosmological starvation, are required for the suppression of star formation in the regions outside the bar region and globally quench the galaxies, if bar quenching is the primary quenching mechanism responsible for transforming barred galaxies into passive galaxies then  we expect galaxies with a longer bar to be more offset from the main sequence in the total star formation rate--stellar mass (SFR--M$\star$) plane. The star-forming galaxies in the local Universe populate the main sequence within an intrinsic scatter of 0.3 dex, whereas the passive galaxies are offset from the relation in the SFR--M$\star$ plane \citep{Brinchmann_2004,Salim_2007,Noeske_2007,Elbaz_2007,Daddi_2007}. We note that such a trend between the scaled bar length (bar length divided by the size of galaxy) and the offset distance from the main sequence was not observed for the barred galaxies in the local Universe (\citealt{George_2019b}, z $\leq$ 0.06), suggesting that the bar-quenching mechanism may not be contributing significantly to the global quenching of barred galaxies.\\

In \citet{George_2019b}, we used the total SFR of the galaxies provided by the  MPA/JHU SDSS DR7 catalogue \footnote{https://wwwmpa.mpa-garching.mpg.de/SDSS/DR7/}\citep{Kauffmann_2003, Brinchmann_2004}, which are derived from the H$\alpha$ flux measured by the 3"  SDSS fibre spectra for star-forming galaxies and using an empirical
relation between specific SFR and the 4000 $\AA$ break strength (D4000 index) for active galactic nuclei (AGNs) or weak-emission-line galaxies. Aperture corrections are done on contribution to SFR outside the SDSS fibre from the spectral energy distribution (SED) fitting to SDSS ugriz photometry. Recently, \citet{Cortese_2020} reported that the 
aperture-corrected SFR estimates from the MPA/JHU SDSS DR7 catalogue could be an underestimate of the total SFR and do not provide a true representation of the global SFR of galaxies with extended star-forming discs. This suggests that any ongoing star formation outside the region covered by the SDSS 3" fibre is not fully accounted for when computing the total SFR of the galaxy. The SDSS 3" fibre does not cover the entire bar region of the sample galaxies (in the entire redshift range) studied by \citet{George_2019b}. \\

This implies that, in \citet{George_2019b}, we were effectively probing the SFR and hence the quenching happening in the very central region of the barred galaxies and not the total SFR of the galaxy nor the SFR of the bar region. This could explain the lack of  correlation between the length of the bar and the offset distance of galaxies from the main sequence of the SFR--M$\star$ relation for galaxies in the local Universe \citep{George_2019b}. Thus, the choice of SFR proxy is important to understand the significance of bar quenching in the local, barred disc galaxies. \\

In this context we aim to study the propagation of star formation quenching in the local barred disc galaxies and their dependence on scaled bar length using total SFR estimates from the GALEX-SDSS-WISE Legacy Catalog (GSWLC) \citep{Salim_2016} and to compare them with estimates from the MPA-JHU SDSS DR7 catalogue \citep{Kauffmann_2003, Brinchmann_2004}.  \cite{Salim_2016} estimated the SFR by fitting SEDs to the global ultraviolet (UV), optical, and infrared\ (IR) flux values of the galaxies. Ultraviolet and mid-infrared(MIR)-based SFR estimates are a better proxy for global SFR of nearby galaxies \citep{Cortese_2020} and hence we use such estimates for checking the dependence of SFR on bar length for a sample galaxies studied in \citet{George_2019b}. In addition, we compare the SFR estimates from the MPA-JHU SDSS DR7 catalogue and GSWLC to differentiate the barred galaxies, which are only centrally quenched, from those that are globally quenched. This in turn helps us to constrain the propagation of star formation quenching  in barred galaxies and can provide more insights into inside-out quenching of star formation in disc galaxies.\\

 We adopt a flat Universe cosmology with $H_{\rm{o}} = 71\,\mathrm{km\,s^{-1}\,Mpc^{-1}}$, $\Omega_{\rm{M}} = 0.27$, $\Omega_{\Lambda} = 0.73$ \citep{Komatsu_2011}.\\

\section{Data and analysis}

\begin{figure*}
\centering
\subcaptionbox{}{\includegraphics[width=0.4\textwidth]{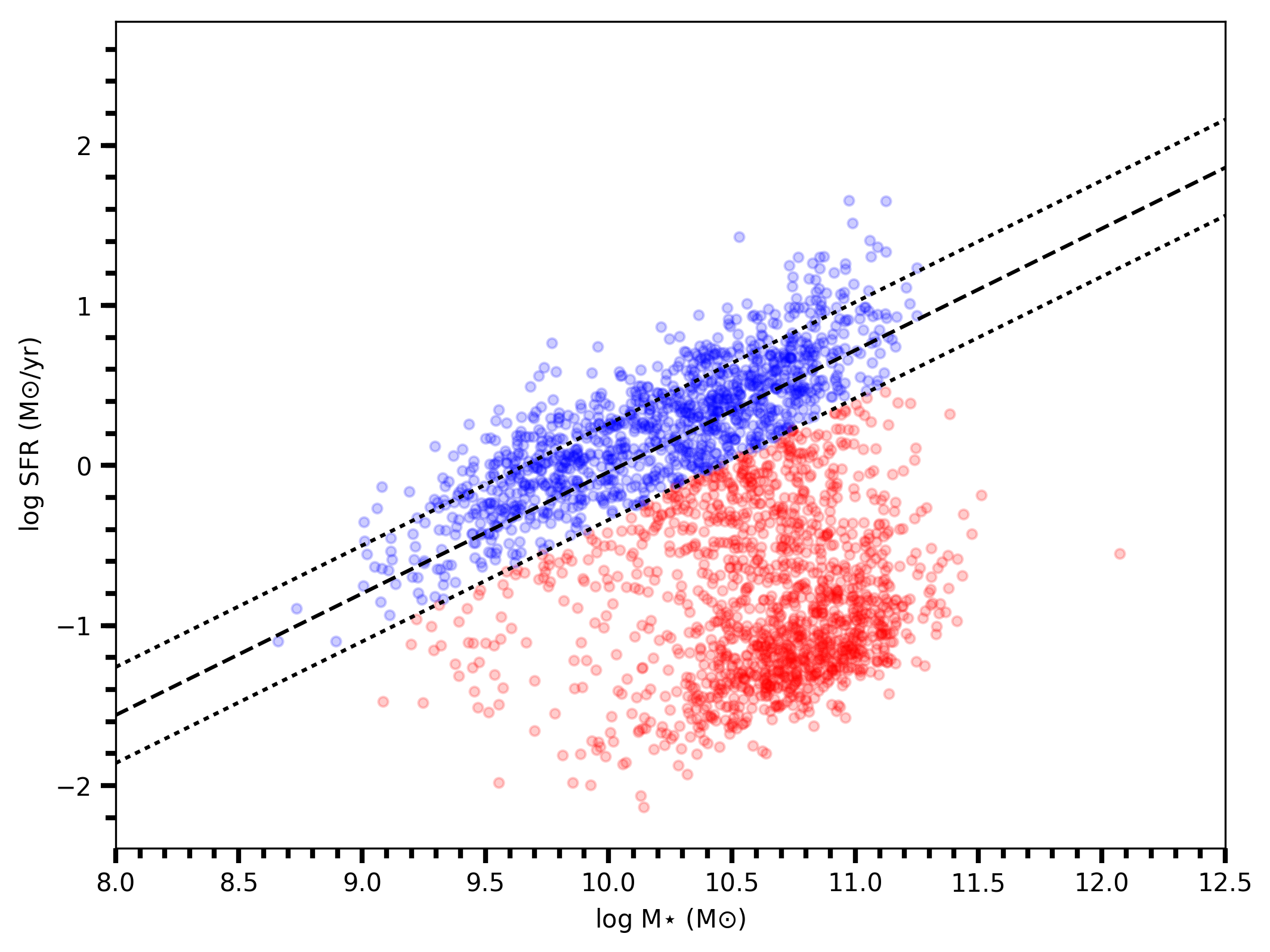}}
\subcaptionbox{}{\includegraphics[width=0.4\textwidth]{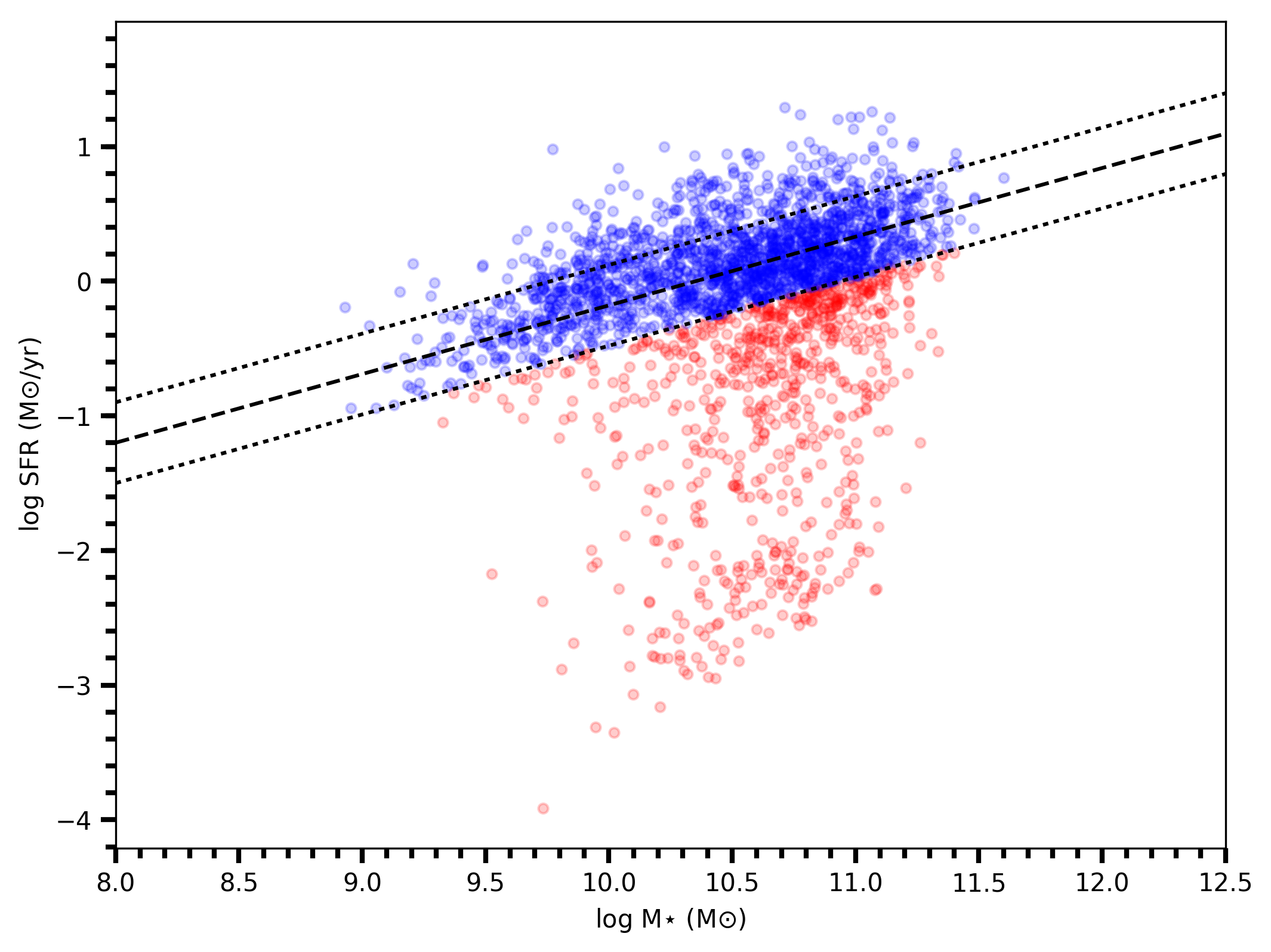}}
\caption{(a) SFR--M$\star$ relation for barred galaxies using data from the MPA-JHU catalogue (b) SFR--M$\star$ relation for barred galaxies using data from GSWLC. The galaxies that are on the main sequence in both SFR--M$\star$ relations are shown in blue and the passive galaxies are shown in red. The main sequence relation is shown with the black dotted line and the 0.3dex width on either side of main sequence is shown with a dashed line. The colour scheme in Fig. \ref{figure:fig2}b is the same as in  Fig. \ref{figure:fig2}a.}\label{figure:fig2} 
\end{figure*}

We followed the same methods as those described in  \citet{George_2019b} using the catalogue of stellar bar length for 3150 barred galaxies. This is produced from a morphological analysis by Galaxy Zoo2 \citep{Willett_2013} and presented in \citet{Hoyle_2011} for face-on galaxies in local Universe with redshifts up to  ($z$) $\sim$ 0.06. We used the scaled bar length of galaxies from this catalogue.\\

 First, the SFR and  stellar mass (M$\star$) of our sample barred galaxies are taken from the MPA-JHU SDSS DR7 catalogue  \citep{Kauffmann_2003, Brinchmann_2004}. As described earlier, the total SFR (corrected for extinction and aperture) in this catalogue is computed from H$\alpha$ flux for star-forming galaxies, and for the case of active galactic nuclei (AGNs) and composite galaxies, the total SFR is derived from $D4000$ index using the method described in \citet{Kauffmann_2003}. The total stellar mass of each galaxy is derived from fitting the broad band $ugriz$ SDSS photometry using the stellar population models of \citet{Bruzual_2003}, assuming a \citet{Kroupa_2001} initial mass function (IMF). Of the 3150 barred sample galaxies, we were able to retrieve reliable non-zero values of parameters for 3068 of them. There can be contamination from higher excitation emission lines due to AGNs at the centre of barred galaxies, which could affect the SFR and M$\star$ estimates. We therefore used the line diagnostic classification based on the emission line kinematics results for SDSS using GANDALF (emissionLinesPort) and  retrieved the BPT classification for 3046 galaxies \citep{Baldwin_1981,Sarzi_2006}. From these, we removed 161 galaxies  classified as "Seyfert" galaxies.  Thus, we have 2885 barred galaxies with reliable SFR and M$\star$ estimates.\\

We cross-matched these 2885 barred galaxies with SFR and M$\star$ estimates from the MPA-JHU catalogue with the GSWLC  \citep{Salim_2016}. GSWLC covers 90$\%$ of the SDSS area and contains galaxies within the GALEX footprint with or without a UV detection.  We used the SFR in GSWLC derived from the broad-band flux of the galaxy used to construct the UV--optical SED followed by a subsequent fit and measures the total flux from the galaxy. We find 2514 galaxies common to both catalogues. We note that these are galaxies with no contamination from AGN emission that could potentially lead to overestimation of the SFR values.\\

We subsequently constructed the SFR--M$\star$ (in log scale) plane for the 2514 barred galaxies from the MPA-JHU catalogue (Fig. \ref{figure:fig2}a). We used the main sequence relation for the local Universe galaxies described in \citet{Renzini_2015}, shown with a black line ($\pm$ 0.3 dex from the relation shown with the dotted line) in Fig. \ref{figure:fig2}a. The galaxies on the main sequence and above (star-forming and star burst galaxies) are marked in blue and those below the 0.3 dex deviation from the relation are marked in red and considered as passive galaxies. The reported scatter for the main sequence relation is 0.3 and therefore our way of distinguishing passive galaxies as not on the main sequence is physically motivated. We also note that we only consider those galaxies that are away from main sequence region as passive, and do not consider the sub-population of green valley galaxies often found in the literature, which are based on positions very near to the main sequence.\\

The SFR--M$\star$ plot made using values from GSWLC for the same 2154 barred galaxies is shown in Fig. \ref{figure:fig2}b. We used the main sequence relation for the local Universe galaxies described in \citet{Guo_2019} and shown with a  black line ($\pm$0.3 dex deviation shown with a dotted line) in Fig. \ref{figure:fig2}b. We see that a good fraction of the barred galaxies that are below the main sequence in Fig. \ref{figure:fig2}a are on the main sequence in Fig. \ref{figure:fig2}b. We find that a good fraction of passive galaxies identified from the SFR--M$\star$ plot using the values in the MPA-JHU catalogue are not passive in the SFR--M$\star$ plot obtained using the estimates from GSWLC.  These might be the galaxies that are centrally quenched but still host star formation in the regions outside of the SDSS 3" fibre. The fraction of galaxies that are passive in both the plots must be globally quenched galaxies. There are galaxies on the main sequence in both the plots and these are the true star-forming galaxies whose star formation is  regulated by their availability of fuel. The total numbers of barred galaxies in different regions of the SFR--M$\star$ plot are given in Table 1. There are 651 galaxies that are passive based on the estimates from the MPA-JHU catalogue but are star-forming based on the estimates from GSWLC.

\begin{table*}
\centering
\label{SFR proxy and bar quenching}
\tabcolsep=0.6cm
\begin{tabular}{ccccc} 
\hline
SFR  & Main sequence & Star burst & Quenched & Total\\
 \hline
GSWLC               & 1468  & 340  & 706  & 2514\\
MPA-JHU             & 886   & 274  & 1354  & 2514\\
\hline
\end{tabular}
\caption{\label{t12} Number of barred galaxies in different regions of SFR-M$\star$ planes constructed using GSWLC and the MPA-JHU catalogue. There are 651 galaxies that are passive in MPA-JHU but are on the main sequence in GSWLC.}
\end{table*}

\begin{figure}
\includegraphics[width=8.5cm]{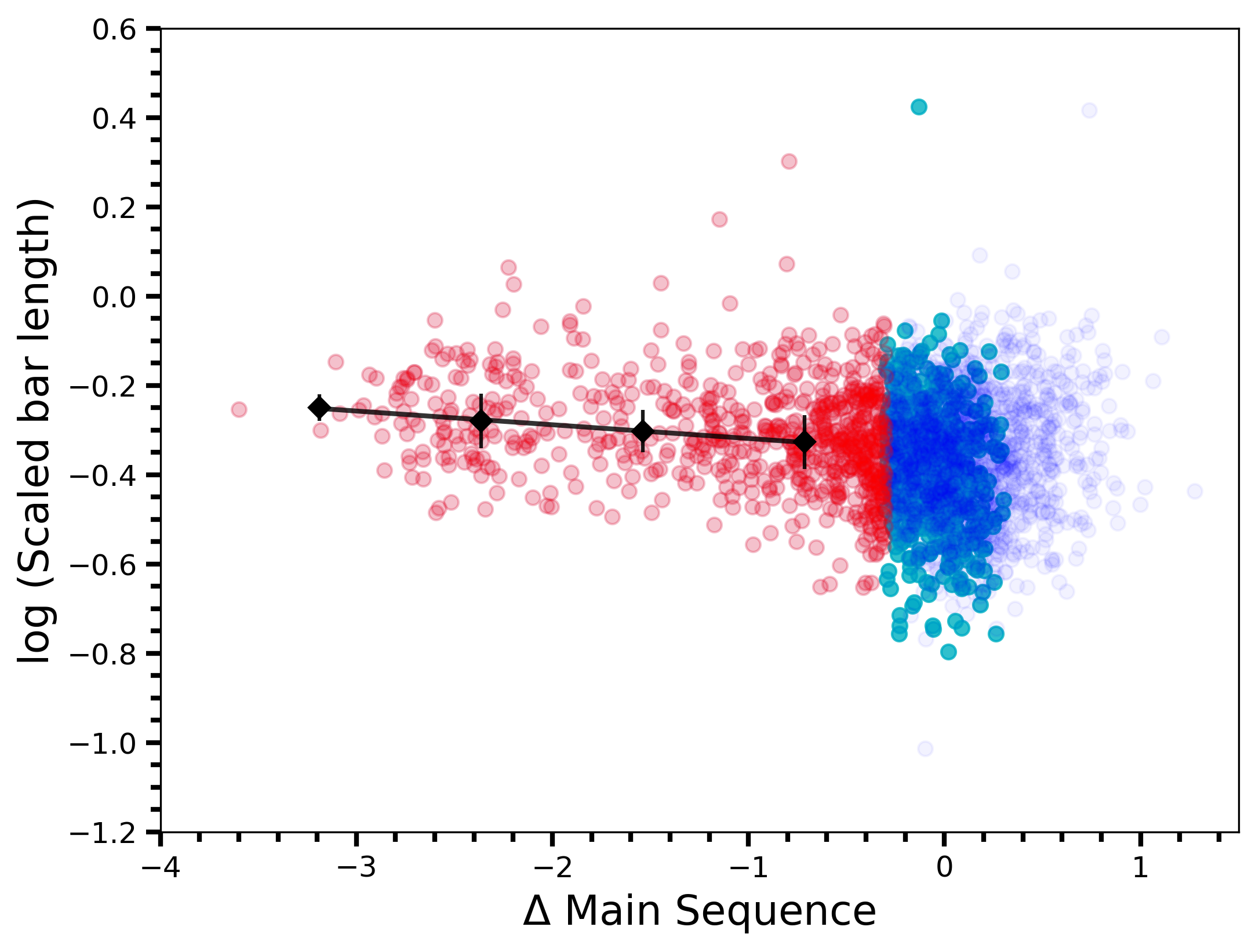}
\caption{Offset of galaxies from the main sequence relation derived from  GSWLC plotted against scaled bar length. The offset of barred galaxies are grouped into four bins so that a possible trend can be investigated. The median value of scaled bar length in each bin is shown with black diamond points with error bars. The best-fit relation for binned points is shown with a black line. The passive population of galaxies 0.3dex below the main sequence is shown in red. Galaxies on the main sequence and 0.3dex above the relation are shown in blue. The 651 galaxies that are passive in MPA-JHU but are on the main sequence in GSWLC are shown in cyan.} \label{figure:fig3}
\end{figure}

\section{Discussion}

The offset of galaxies from the main sequence relation of SFR--M$\star$ created from GSWLC are plotted against the scaled bar length in Fig \ref{figure:fig3}. There exists a  correlation between the offset distance of passive galaxies from the main sequence in the SFR--M$\star$ plot constructed using the values from GSWLC and the scaled bar length. These galaxies that are away from the main sequence relation might be globally quenched, primarily due to the action of the stellar bar. The galaxies on the main sequence show no trend with scaled bar length, but  as passive galaxies move away from the main sequence they begin to show a mild dependence on scaled bar length, that is, both individual galaxies and the median binned value of galaxies in the plot.  We note that such a relation was not seen in \citet{George_2019b} where the MPA-JHU catalogue was used to create the SFR--M$\star$ plot.\\

The best-fit relation for the binned points is shown with a black line. We note that the choice of bin size and number of bins does not affect the best-fit relation. We derived best-fit parameters by changing the number of bins; using 3, 4, 5, or 10 bins gives very similar results. We parameterize this dependence as follows: \\

\begin{equation}
\Delta MS = -0.03 log(L_{bar}) - 0.35\\
\end{equation}

where $\Delta MS$ is the offset of galaxies from the main sequence relation and $L_{bar}$ is the scaled bar length. We used the best-fit SFR--M$\star$ relation from \citet{Guo_2019} for the galaxies in GSWLC and substituting for $\Delta MS$ = logSFR$_{observed}$ - logSFR$_{predicted}$ we found a relation between scaled bar length and SFR shown as follows.\\

\begin{equation}
logSFR = 0.51 logM_{\star} - 0.03 log(L_{bar}) - 5.63\\
\end{equation}

 The observed weak dependence in  passive galaxies between bar length and offset  from the main sequence relation could be due to the role of bars in enhancing the central star formation. Stellar bars can initially enhance the star formation in the central sub-kpc region of barred galaxies. Recently, \citet{Lin_2020} reported that, for barred  galaxies, the radius of the central `turnover region' in H$\alpha$ and H$\beta$ equivalent width (indicative of enhanced recent star formation) correlates with bar length. Therefore, even when the action of the stellar bar quenches star formation in the bar region and acts as a primary mechanism for global quenching of star formation in the barred galaxies, the presence of any residual flux in the central region can weaken the expected  correlation between bar length and global SFR.

\subsection{Dependence on stellar mass}

In this section, we check for any dependence of galaxy stellar mass on the relation between the offset of passive galaxies from main sequence and scaled bar length. We divided the sample into mass bins, M$\star$ $>=$ 10$^{10.5}$ and  M$\star$ $<=$ 10$^{10.5}$. There are 502 quenched galaxies out of 1491 galaxies with M$\star$ $>=$ 10$^{10.5}$ and 204 quenched galaxies out of 1023  galaxies with M$\star$ $<$ 10$^{10.5}$. We derived the best-fit relation between scaled bar length and offset from main sequence for the binned points as in previous section.\\

\begin{equation}
M_{\star} < 10^{10.5};
\Delta MS = -0.03 log(L_{bar}) - 0.32\\
\end{equation}

\begin{equation}
M_{\star} >= 10^{10.5};
\Delta MS = -0.02 log(L_{bar}) - 0.35\\
\end{equation}

The relation shows a slight but insignificant deviation at the high-mass end.

\subsection{Environmental dependence}

\begin{figure}
\includegraphics[width=8.5cm]{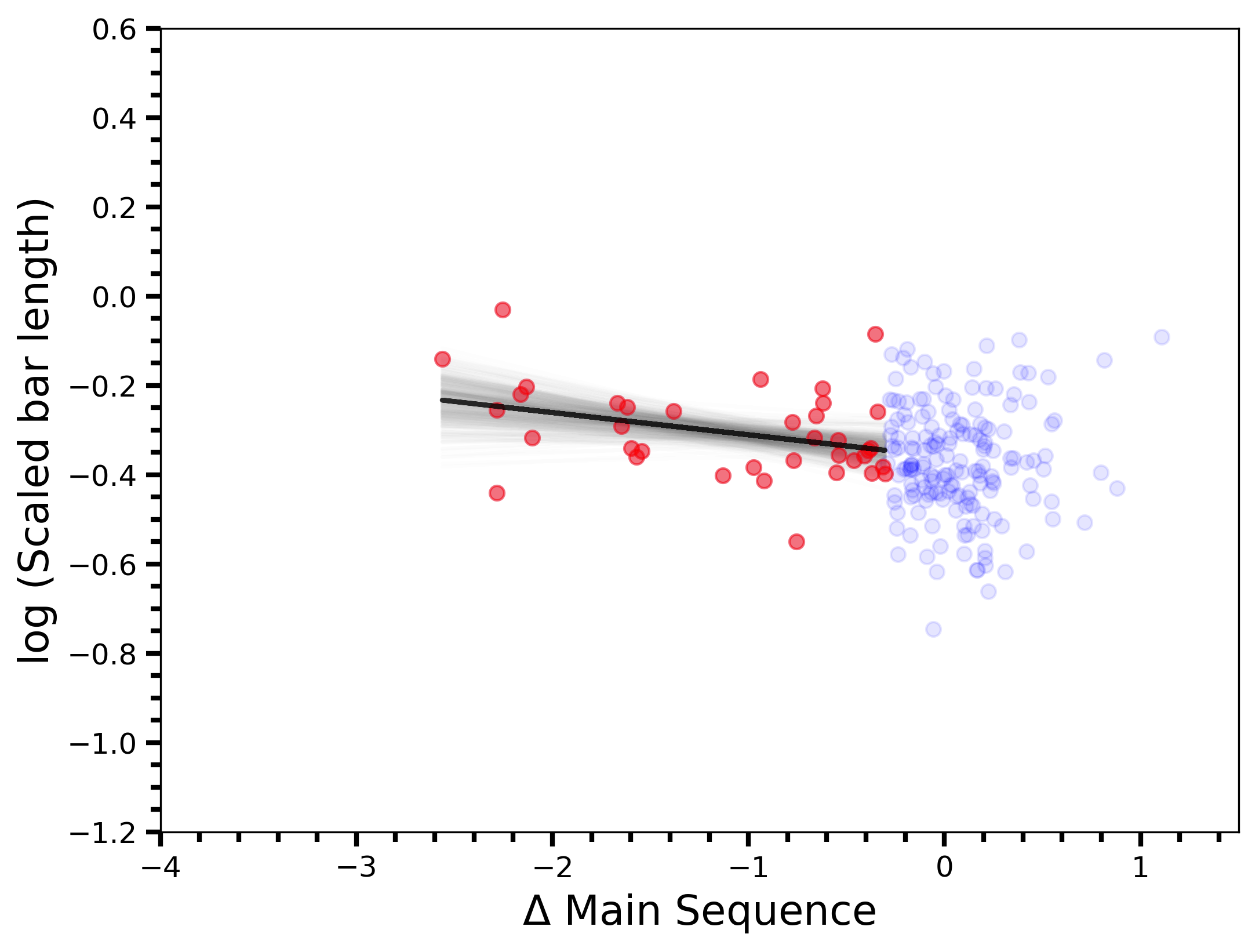}
\caption{Offset of 210 isolated barred galaxies from the main sequence relation derived using GSWLC plotted against scaled bar length. Here the number of data points is limited and hence instead of binning the data we bootstrapped the data with confidence intervals shown in grey and the best-fit relation  shown in black. The passive population of galaxies  below the main sequence is shown in red. Galaxies on and above the relation are shown in blue.} \label{figure:fig4}
\end{figure}

The local environment can affect the evolution of galaxies, and for barred galaxies the presence of a companion can influence the length of the bar and star formation. We checked for possible dependencies of the relation between offset of passive galaxies from the main sequence and scaled bar length on environmental factors. We used the catalogue of isolated galaxies in the local Universe from SDSS below redshift 0.08 compiled in \citet{Argudo_2015}. We find that out of 2514 barred galaxies in our sample, 210 galaxies are classified as isolated with no influence from companions during the past 3 Gyr. There are 173 isolated galaxies on the main sequence and 37 galaxies in the passive region of the SFR--M$\star$ plane made with GSWLC as shown in Fig \ref{figure:fig4}. The sample size of isolated galaxies is limited compared to the total galaxy sample used in Fig. 2. We therefore followed a bootstrap procedure to derive the best-fit relation between scaled bar length and offset from main sequence for the binned points. The best-fit relation is given below.

\begin{equation}
\Delta MS = -0.05 log(L_{bar}) - 0.36
\end{equation}

We find that the  dependence between the offset of barred galaxies from the star-forming MS and scaled bar length is marginally higher for isolated galaxies than for the entire sample. The nature of star formation progression in these galaxies needs to be checked, possibly with a detailed emission line kinematic analysis using integral field unit(IFU)-based observations. \\

The sample of 651 barred galaxies that are centrally quenched could be undergoing inside-out quenching where the action of the stellar bar suppresses star formation in the very central region. A detailed, spatially resolved stellar population in conjunction with gaseous and stellar kinematic studies using  IFU data of barred galaxies could provide further details of the processes happening in these galaxies that are associated with bar quenching (see \citealt{Krishnarao_2020}).\\

Stellar bars are formed in a galaxy when it is dynamically cold and  stable. Therefore, at high redshifts, galaxies do not favour stellar bar formation, where frequent galaxy mergers and  stellar feedback leads to dynamically hot systems where bar formation cannot happen. Recently, observational evidence was found  for the existence of rotating, dynamically cold, star-forming  discs   ---very similar to the dynamics of local Universe galaxies--- at redshifts of $\sim$ 4 \citep{Rizzo_2020}. Also, a large fraction of recently quenched green valley galaxies were found to have quiescent central regions, while hosting star formation in extended outer discs \citep{Belfiore_2017}. If stellar bars are formed at high redshifts then bar quenching could be a dominant quenching channel, moving galaxies from the star-forming main sequence to the passive population of galaxies.\\

\section{Summary}
We classified barred galaxies as centrally quenched and globally quenched based on their position on SFR--M$\star$  plots, with SFR derived from H$\alpha$ flux and ultraviolet and optical flux. We selected galaxies as quenched in the central regions if they fall in the passive region of the SFR--M$\star$  plot created using the MPA-JHU catalogue and in the main sequence region of the same plot created using GSWLC. This difference in sampling is due to the fact that the finite size of the SDSS fibre used for measuring galaxy SFR in the MPA-JHU catalogue samples star formation in the very central regions of the galaxy. In the present paper, we present a sample of 651 barred galaxies with suppressed star formation in the central region but hosting star formation outside. We also find a possible correlation between bar length and SFR for the globally quenched galaxies and suggest that bar quenching may be contributing significantly to the global quenching of star formation in barred disc galaxies. This correlation become slightly stronger when only isolated barred galaxies are considered.

\begin{acknowledgements}
 SS acknowledges support from the Science and Engineering Research Board of India through a Ramanujan Fellowship. Funding for the SDSS and SDSS-II has been provided by the Alfred P. Sloan Foundation, the Participating Institutions, the National Science Foundation, the U.S. Department of Energy, the National Aeronautics and Space Administration, the Japanese Monbukagakusho, the Max Planck Society, and the Higher Education Funding Council for England. The SDSS Web Site is http://www.sdss.org/. The SDSS is managed by the Astrophysical Research Consortium for the Participating Institutions. The Participating Institutions are the American Museum of Natural History, Astrophysical Institute Potsdam, University of Basel, University of Cambridge, Case Western Reserve University, University of Chicago, Drexel University, Fermilab, the Institute for Advanced Study, the Japan Participation Group, Johns Hopkins University, the Joint Institute for Nuclear Astrophysics, the Kavli Institute for Particle Astrophysics and Cosmology, the Korean Scientist Group, the Chinese Academy of Sciences (LAMOST), Los Alamos National Laboratory, the Max-Planck-Institute for Astronomy (MPIA), the Max-Planck-Institute for Astrophysics (MPA), New Mexico State University, Ohio State University, University of Pittsburgh, University of Portsmouth, Princeton University, the United States Naval Observatory, and the University of Washington. Finally, it is a pleasure to thank the anonymous referee for providing useful and encouraging comments.
\end{acknowledgements}


\end{document}